# Autonomous animal heating and cooling system for temperature-regulated MR experiments

George Verghese[1,2], Mihaly Voroslakos[3], Stefan Markovic[4], Assaf Tal[4], Seena Dehkharghani[1,2], Omid Yaghmazadeh[3*], Leeor Alon[1,2*,]

[1]*Center for Advanced Imaging Innovation and Research (CAI²R), New York University School of Medicine, New York, NY, United States,* [2]*Center for Biomedical Imaging, New York University School of Medicine, New York, NY, United States.* [3]*Neuroscience Institute, NYU School of medicine, New York, NY.* [4]*Department of Chemical Physics, Weizmann Institute of Science, Rehovot, Israel.*

**Key Words:** Animal temperature, MR thermometry, preclinical MRI, animal thermoregulation.

**Abstract**

Temperature is a hallmark parameter influencing almost all magnetic resonance properties (e.g., $T_1$, $T_2$, proton density, diffusion and more). In the pre-clinical setting, temperature has a large influence on animal physiology (e.g., respiration rate, heart rate, metabolism, cellular stress, and more) and needs to be carefully regulated, especially when the animal is under anesthesia and thermoregulation is disrupted. We present an open-source heating and cooling system capable of stabilizing the temperature of the animal. The system was designed using Peltier modules capable of heating or cooling a circulating water bath with active temperature feedback. Feedback was obtained using a commercial thermistor, placed in the animal rectum, and a proportional–integral–derivative (PID) controller capable of locking the temperature. Operation was demonstrated in a phantom as well as mouse and rat animal models, where the standard deviation of the temperature of the animal upon convergence was less than a tenth of a degree. An application where brain temperature of a mouse was modulated was demonstrated using an invasive optical probe and non-invasive magnetic resonance spectroscopic thermometry measurements.

**Introduction**

Preclinical animal Magnetic Resonance Imaging (MRI) is a vital tool for studying biological processes in vivo[1]. In recent years, a growing number of preclinical MRI studies have underscored the multi-contrast capabilities of MRI to inform on tissue physiology, anatomy, and function. Often, preclinical imaging studies require anesthetizing animals to minimize stress and reduce motion artifacts. Under anesthesia, however, temperature regulation is disturbed[2] and gradually drops[3–5], thus special care must be taken to maintain the animal's body temperature[5,6] as it may affect the animal's wellbeing and minimize the risks from the anesthesia itself. The anesthesia increases heat loss due to the inhalation of cold gasses and the disruption of the animal's normal thermoregulation and behaviors for regulating temperature. This heat loss is exacerbated given the rodent's large surface-to-body volume ratio, which without assisted devices used to regulate the animal's temperature, would lead to hypothermia and death[5]. Temperature has been shown to modulate the animal's physiology affecting the respiration rate and depth[7,8], cardiac function[9], brain activity[10], and metabolism[7], all of which could influence MR experiments. These physiological features are especially important in investigations of biological function[11,12].

Temperature affects not only physiological parameters, but also image contrast, with longitudinal relaxation ($T_1$)[13] exhibiting a linear temperature dependence of ~1.4%/ºC in bovine and rabbit muscle[14,15], 1-2%/ºC in ex vivo liver[16], and 0.97%/ºC in fat[17]. Transverse relaxation ($T_2$) exhibits a reduction of ~6%/ºC with increase in temperature in fat[18], while proton density and the diffusion constant depend linearly with temperature changing at a rate of ~0.3%/ºC and ~2-2.5%, respectively, in many tissues [19–21]. In addition, temperature-related pH and magnetization transfer changes have also been reported[22–24]. Temperature is responsible for changing the electrical shielding around proton at a rate of ~-0.01 parts per million (ppm)/ºC[25–27], with larger changes for other nuclei such as sodium, xenon, and phosphorous[28–30]. This is accompanied by temperature-dependent magnetic susceptibility changes of ~0.003 ppm/ºC[26,31,32]. Further alterations to tissue electrical properties (i.e., conductivity and relative permittivity) with temperature have been reported[33], which can affect RF transmit and receive magnetic fields throughout the time course of the MR experiment and cause minor alterations to image contrast (by changing the propagation of electromagnetic fields in the body), especially at ultra-high magnetic field strength. Therefore, almost all electric and magnetic properties of tissues are, in fact, temperature dependent[33].

The influence of temperature change on MR tissue contrast and animal physiology thus compel development of tools to minimize, or ideally eliminate, their effects during in vivo experimentation and improve quantitative MR. Commercial MR-compatible heating systems for animals often utilize an external resistive heating source with air or a water transport system used to heat a reservoir of water that is then pumped adjacent to the animal[34]. Through conduction, the heated water reservoir transfers heat to the animal such that the animal body temperature is maintained. One study demonstrated heating the animal using a resistive heating element placed under the animal, where currents were driven at 10-100 KHz to mitigate interaction with the static magnetic field that can remain beyond the perceptible auditory range of rodents[35]. Animal heating, is carried out often in conjunction with contemporaneous temperature measurements of the animal, typically accomplished by placement of a temperature probe inside the rectum of the animal. In most cases, manual adjustments are made throughout the experiment to maintain the temperature within a certain threshold; however, this process is highly user dependent. Importantly, proprietary MR-compatible animal heating systems do not provide cooling capabilities necessary to apply a precise and fast control of the core temperature, which might increase as a consequence of the utilization of high-power sequences that deposit energy in tissue or may be required in cooling protocols such as those employed in the study of therapeutic hypothermia for cerebrovascular or other injury[36]. In addition, these systems do not offer real-time control to maintain the temperature of the animal without intervention. Such control is highly desirable when conducting MR thermometry experiments or functional MRI studies, where thermal changes in time might yield ambiguous results[37].

The goal of this work was to introduce an open-source system with feedback for tightly controlling animal temperature in a cost-effective manner (<$500 in materials), which was relatively easily constructed and would improve quantitative in vivo MR experimentation. The system for heating and cooling of small animals utilized Peltier modules[38,39] that controlled the temperature of water within a small circulating water reservoir. Water was transported to a custom-made 3D-printed plastic cradle with water channels capable of conductively heating or cooling the animal. The system was capable of both setting, monitoring and autonomously maintaining the body temperature of the animal using feedback provided by a thermistor which was made RF safe for use in the MR environment. Temperature control was achieved using a proportional–integral–derivative (PID) controller[40,41], employing a feedback control loop mechanism to continuously modulate the thermal output of the system such to ensure body temperature stability. Operation was demonstrated in phantom as well as mouse and rat in vivo models, where the standard deviation of the temperature of the animal upon convergence was less than a tenth of a degree. An application where brain temperature of a mouse was modulated was demonstrated using an invasive optical probe and non-invasive magnetic resonance spectroscopic thermometry measurements.

**Methods**

*Electronics:* An Arduino UNO R3 (EL-CB-001, Elegoo) microcontroller using an ATmega328 microcontroller governed the autonomous attribute of the heating and cooling system. The heating and cooling mechanisms were achieved by the use of four standard TEC1-12708 Peltier modules. A pair of the Peltier modules were wired in series, where for each series connection, a single H-bridge (Infineon BTS7960 chip) was used to drive the module pair by switching the polarity of the voltage to the Peltier modules and thus switching the respective cooling or heating surface. For each H-bridge, the M+ and M- terminals were connected to their respective pairs on the Peltier modules and driven from a common +15V rail. Each pair of TEC1-12708 Peltier module consumed approximately 3.5Amps, such that the total current drawn for the Peltier modules was approximately 7Amps in heating or cooling modes. To facilitate these currents, 20 American Wire Gauge (AWG) wire was used for all Peltier connections, while 26AWG was used for all other connections. The heating and cooling modes were designed in software on the Arduino UNO and pulse width modulation (PWM) was used to control the temperature, where a zero PWM value corresponded to zero duty cycle and 255 to one hundred percent duty cycle. M+ and M- of the H bridge 1 were also connected to a 5mm Red LED (D1) and a 5mm Blue LED (D2), wired in antiparallel configuration through a 10kΩ resistor (R4) to indicate when the system is in the heating (Red LED ON) or cooling mode (Blue LED ON). The entire system used four discrete positive voltages (+15V, +12V, +8V, and +5V) to drive the various subassemblies from a 24V Switching Power Supply (DROK-200571). The highest voltage of +15V was used to drive the H-bridges directly from the main DROK power supply and was fed to two three-terminal positive regulators (L78S12CV, STMicroelectronics and MC7808CTG, Onsemi), to derive the +12V and +8V voltage rails, respectively. The L78S12CV with the associated decoupling capacitors was used to power the four fans connected in parallel and mounted on a heat sink, as well as the Water Pump located within the main electronics box enclosure. The main electronics box enclosure was placed at the farthest end of the MR table of the Bruker system, to prevent any interference with the MR system electronics. The +8V rail derived from the MC7808CTG was fed via a toggle switch to the Arduino UNO R3 via its DC input barrel jack, and a voltage regulator (AMS1117-5, Advanced Monolithic Systems) housed on-board the Arduino UNO R3, steps down the +8V to +5V. Two safety fuses were used in the design: the first, rated for 2Amps and in line with the AC mains (on the live wire) before being fed to the power supply and a second 15Amps slow blow ceramic fuse, was used in line with the +15V output of the power supply to protect the system in case any normal working currents were exceeded. A combined voltmeter/ammeter module, as shown in Figure 1A, monitors both the system voltage and current where the voltmeter input was tied directly to the +15V rail of the power supply, while the shunt in the ammeter was connected in series to monitor the current draw.

In the experiments, two Negative Temperature Coefficient (NTC) thermistors were used. Epoxy sealed NTC1 (NRL1104F3950B1F, Bussmann / Eaton) with a Beta value of 3950 and a nominal resistance of 10kΩ at 25ºC, formed a potential voltage divider with a fixed 10kΩ metal film resistor. NTC1 was galvanically connected to a 350cm twisted pair wire made of American Wire Gauge 26 (AWG26) that was inserted rectally into the rodent being imaged. Common mode currents which were induced on the twisted pair wire from the transmit field of the MR system was decoupled using four 1uH chokes (1812CS-102XJR, Coilcraft) at λ/4 distances (where λ=~25cm in free space at a Larmor frequency of 298MHz for a 7T MRI scanner). In order to ensure that the rodent bed does not exceed 45ºC, an interlock was designed using a thermistor NTC2 (B57702M0103J000, TDK) which was mechanically mounted on one of the heat exchanger's and utilized similar voltage divider circuitry to that used for NTC1, in order to estimate the temperature of the water bath. When the temperature exceeded 45ºC, the H-bridge was pulled LOW, thus cutting off the power to the Peltier modules. Target temperature for the rodent core-body temperature was set using a potentiometer (WTH118-2W, Phoncoo), which was located on the front of the enclosure as shown in Figure 1B. The target temperature and the current temperature of the rodent was displayed on a 16x2 I$^2$C, controlled LCD Display Module (part I2C 1602 LCD Display Module, GeeekPi) and powered by the Arduino UNO. A 4.7kΩ metal film resistor (R3) was connected in series with a 1kΩ trimmer resistor (PV36W102C01B00, Bourns Inc.) to adjust the resistance of the series combination to exactly 5kΩ. The node connecting the 1kΩ trimmer and the potentiometer was fed to the analog-to-digital (A/D) converter pin A2 on the Arduino UNO, and the voltage converted to the corresponding temperature, was displayed as the target temperature on the LCD display. A temperature range between 15ºC and 45ºC was readily achieved using this setup. Cost of materials were under $500 and the materials and equipment needed was readily available through commercial vendors.

*Bed design*: The heating and cooling bed was designed using the Fusion 360 software (Autodesk Inc., California, USA) with the following dimensions: width=36 mm, length = 200 mm, and height = 5.6 mm (Figure 1C), such that the bed was compatible with our previously reported open-source cradle for animal imaging[42]. The bed included 4 -traversing water channels, 4.6 mm in diameter, going through length of the bed (Figure 1C, bottom), such that water can flow throughout the bed and create a uniform temperature distribution capable of transferring or taking away heat to/from the animal. At one extremity of the bed, two 3/16" in diameter and 17.65 mm in length, barbed water-pipe connectors were designed, in order to connect 3/16" ID silicon tubing from the electronics box to the bed. The bed was printed using a Formlabs Form 3B (Formlabs Inc., Massachusetts, USA) stereolithography (SLA) 3D printer using their biomed clear resin. Upon completion of the printing process, the bed was washed and UV-cured, and connected to the main electronics box with silicon tubing allowing water to be circulated from the Peltier heat exchangers to the bed. The water within the water bath was doped with Gadobutrol to shorten the $T_1$ relaxation such that the water bath is MR-invisible while scanning an animal with the water bed in place. The 3D design files and the bill of materials are available online, and the link to the materials can be found in the supplemental materials section.

*Temperature feedback and control:* Temperature control was achieved using a PID controller, a commonly used methodology to control industrial control systems and a variety of other applications. The PID controller was implemented using the Arduino PID library[43], where the temperature measured from the animal and the target temperature were used as inputs. Once per second, the error between the current temperature and target temperature was computed and three terms (P, I, and D) were used to generate a Pulse Width Modulation (PWM) output between 0 and 100% duty cycle to bias the Peltier modules such that heating or cooling was possible at different power levels. The P term was used to scale the error value between the current and target temperature such that the PWM decreases its duty cycle as the current temperature approaches the target temperature. The I term was used to reduce steady state temperature errors, while the D term was used to compute the future trend and to reduce error fluctuations in time. Tuning of the PID parameters was conducted empirically as described here[44], arriving at tuned parameters: $K_p$=320, $K_i$=100, and $K_d$=2 and $K_p$=200, $K_i$=10, and $K_d$=5 for rats and mice, respectively, while for other species, these parameters can further be optimized. Upon completion of the tuning procedure, two heating and cooling experiments were conducted in a water phantom with and without the PID controller with the following sequence of prescribed temperatures at 21 ºC room temperature: 15 ºC, 30 ºC, 45 ºC, 21 ºC, and 37 ºC. The time-course of temperature for these two experiments was recorded using an external optical temperature measurement system (Pico-M, Opsens Inc., Quebec, Canada). Temperature variation was plotted and stability was computed for results with and without PID controller, and infrared thermal images were acquired using a thermal camera (FLIR ONE, FLIR Systems, Inc., Wilsonville, OR) at 15 ºC, 30 ºC and 45 ºC, illustrating temperature uniformity of the bed (Figure 2B). Temperature was also verified using two temperature measurement systems: the thermistor and an optical temperature probe (OTP-M, OpSens Solutions Inc., Canada) to provide ground truth measurement.

*Animals:* For the in vivo experiments, we used mice and a rat that were on waitlist of euthanasia (due to old age or non-positive outcome of breeding of genetically modified animal models). In this way, we avoided purchase and involvement of additional lab animals. For experiment involving mice, adult wild-type mice were used. Mice were kept in cages in a 12hr regular cycle vivarium room dedicated to mice, in up to five-occupancy cages. The rat was kept in a cage in a 12hr regular cycle vivarium room dedicated to rats in up to two-occupancy cages. No prior experimentation had been performed on the animals. All experiments were conducted in accordance with the Institutional Animal Care and Use Committee (IACUC) of New York University Medical Center.

*Animal temperature modulation experiments:* To test the heating and cooling operation alongside the feedback mechanism, experiments were conducted on an anesthetized mouse and an anesthetized rat. A female mouse weighing 25 grams and a male rat weighing 801 grams were used in the experiments. Each animal, independently, was sedated using 2% isoflurane and was allowed to cool down due to anesthesia. Then the target core body temperature was set to two different levels: 36 ºC and 39 ºC., while animal's rectal temperature was measured using the thermistor and controlled using the PID controller. An optical temperature probe (OTP-M, OpSens Solutions Inc., Canada) was added in order to provide a ground truth measurement of the body temperature. Upon completion of the experiments, temperature recordings were extracted and plotted using MATLAB (Mathworks Inc., Natick, Massachusetts, USA). Animals were humanely euthanized after the experiment.

*MR Imaging experiments in phantom:* MR phantom experiments were conducted for assessing signal-to-noise ratio (SNR) and changes that might occur to transmit magnetic field ($B_1^+$) due to the presence of the heated bed system and the thermistor, and also to verify that no RF heating at the thermistor tip was present. To assess the effect of the heating system on the scanner, SNR was measured in a 50cc Falcon tube filled with Phosphate-buffered saline (PBS) solution, where the thermistor was placed at the end of the phantom submerged inside the liquid. The setup was then placed on the water bed and placed inside an 86 mm diameter proton birdcage coil in our Bruker 7T animal scanner. Tuning and matching of the birdcage coil then took place and the bed electronics were turned on. A low flip angle gradient echo acquisition was then run in axial, sagittal and coronal slices across a large field of view (FOV). The sequence was run with the following parameters: TE=4ms, TR=2500ms, flip angle=30º, acquisition time=160 seconds, FOV=80x80 mm$^2$, and matrix size =64x64. Using the same sequence parameters, the flip angle was set to zero and a noise-only image was acquired. Upon completion of the gradient echo acquisitions, the thermistor was removed followed by retuning and matching of the birdcage and the same gradient echo low flip angle and noise acquisitions. Upon completion of these acquisitions, the images were reconstructed and exported to MATLAB for analysis. The SNR was then reconstructed according to [45] and the results were plotted. Similarly, $B_1^+$ mapping with and without the temperature probe was conducted using the double angle method (DAM)[46]. Two gradient echo sequences with a flip angle of 30° and 60° were used, respectively, with a TR of 2500ms. $B_1^+$ maps were reconstructed and plotted. In order to ensure that RF heating at the tip of the probe was not present, the optical thermal probe was tethered to the thermistor and placed inside the phantom, providing temperature information for tip heating. Then, a high specific absorption rate (SAR) $T_2$-weighted anatomical Rapid Acquisition with Relaxation Enhancement (RARE) sequence with a RARE factor set to 1 and 180° refocusing pulses outputting 81.2 Watts of power, was run while recording the temperature at the probe tip over the course of an 80-second scan. This was used to indicate if RF heating of the probe was present prior to animal experiments. Upon completion of the RF heating experiments, temperature recordings were analyzed.

*In vivo MR validation:* Prior to MR experiments, a 26.2-gram male mouse was injected with urethane dose of 1.3-1.5g/kg into the intraperitoneal cavity. Then, using a stereotaxic system, an optical temperature probe (0.3 mm in diameter, PRB-100-01M-STM, Osensa Innovation Corp., BC, Canada) was implanted in the animal's brain through a craniotomy opening of the skull. A second identical optical temperature probe was attached to the thermistor that provides feedback to the heating system for closed-loop temperature control and they were inserted in the animal's rectum. Measurements were acquired using an FTX-300-LUX+ system (Osensa Innovations Corp, BC, Canada). The animal was then placed on the heating/cooling bed while in the scanner bore and body temperature was adjusted to 36 ºC, 34.1 ºC, 36.5 ºC, 38 ºC and 35 ºC. These values were chosen as they sit within safe temperature range for the animal under anesthesia as validated in our group's previous studies[47,48]. The rate at which brain temperature followed body temperatures was recorded in order to infer the estimated brain temperature from the measured body temperature. This was required, as subsequent acquisitions on animals with brain implants was not possible as those reduce the homogeneity of the magnetic field in the brain.

In a different scan session, an intact (i.e., without implanted brain temperature probe) 26.8-gram male mouse was injected with urethane dose of 1.3-1.5g/kg into the intraperitoneal cavity. A 20 mm local surface head coil (Bruker, Billerica, MA) was placed on top of the animal's head and the animal was then positioned inside the 7T scanner (Bruker., Billerica, MA) with an 86 mm transmit-only birdcage used for excitation. The animal temperature was then set to: 36ºC and 37 ºC., respectively, where at each temperature the body temperature of the animal was allowed to stabilize prior to imaging. Upon stabilization at each temperature, shimming was conducted to correct for temperature related susceptibility changes[26,32] followed by proton Point-RESolved Spectroscopy (PRESS) sequence with the following parameters: TE=40 ms, TR=2500 ms, flip angle=90º, averages=150, voxel size=3x3x3 mm$^3$, acquisition bandwidth=6010 Hz, and acquisition time=375 seconds. Proton Magnetic Resonance

Spectroscopy (MRS) thermometry was used as it noninvasively measures temperature based on frequency differences between temperature-dependent water and temperature independent metabolites. The three frequencies of the N-acetyl-aspartate (NAA), creatine (Cr), and choline (Ch) singlets at 2.01, 3.03 and 3.19 ppm were assumed temperature independent and fixed, and used as references to improve the reproducibility of temperature measurement[49]. Upon completion of the acquisitions the data was exported to the TopSpin 4.2 software (Bruker, Billerica, MA), where the frequency difference between water and the metabolites was reconstructed and the average temperature was computed according to:

$$T_{NAA} = 315.6 - 103.8(\delta_{H_2O} - \delta_{NAA})$$

$$T_{Cr} = 206.7 - 101.7(\delta_{H_2O} - \delta_{Cr})$$

$$T_{Ch} = 193.4271 - 106.08(\delta_{H_2O} - \delta_{Ch})$$

$$T_{avg} = (T_{NAA} + T_{Cr} + T_{Ch})/3$$

where $\delta_{H_2O}, \delta_{Cr}, \delta_{Ch}$, and $\delta_{NAA}$ are the frequencies of the water, creatine, choline and N-acetyl-aspartate peaks respectively, in units of ppm. Peak frequencies were identified by recording the frequency at the maximum of the peak magnitude. The reconstructed average temperature of the brain was compared to the set temperature measured by the heating system for non-invasive modulation of brain temperature using this system.

## Results

The complete architecture of the heating box is illustrated in Figure 1A. PID control system was implemented in software and uploaded onto the Arduino UNO microcontroller, allowing flexibility of implementation. Realization of the heating box (front and back) and a 3D view of the animal bed are illustrated in Figure 1B-C. When initial experiments in liquid phantoms were conducted without a PID, the maximum swing in temperature was 0.9 °C (Figure 2A, red line) and the PID implementation was realized in software in order to improve locking onto the target temperature. The PID implementation reduced the maximum swing in temperatures in the phantom to 0.2 °C, as show in Figure 2A, black. A summary table comparing the temperature fluctuations with and without PID can be found in table SM1. Infrared (IR) temperature of the bed at 15 °C, 30 °C, and 45 °C is illustrated in Figure 2B, demonstrating the capability to modulate the temperature of the bed from room temperature (~21°C). The capacity of attaining higher and lower temperatures relative to the animal's temperature is important to ensure steady state temperature when the physiology or environment is changing.

Comparison of $B_1^+$ and SNR with and without the heating system turned on and the thermistor placed inside the phantom is illustrated in Figure 3A and 3B, respectively. Comparison of axial, coronal and sagittal slices demonstrate a $B_1^+$ difference of <5% by placement of the thermistor probe inside the phantom. Since originally this thermistor probe was a commercially available product, a susceptibility artifact leading to a null was observed over a distance of around 11 mm around the location of the probe. SNR maps indicate a similar effect where the placement of the probe inside the phantom minimally influences SNR. Therefore, 15 mm away from the thermistor, standard imaging can be conducted given the minimal interaction between the birdcage fields and the thermistor temperature probe, which was mitigated by placement of 4 cable traps onto the cable. Rectal probe temperature measured using an RF-interference-free optical temperature measurement system while applying the high-SAR RARE sequence is shown in Figure SM 1, confirming the mitigation of RF-interactions between the thermistor and the birdcage leading to no temperature change on the probe across an 80-

second scan. Confirmation of the stable temperature of the thermistor was further confirmed in in vivo experiments.

Feasibility of the system on mouse and rat animal models is shown in Figure 4. For mouse and rat models, different PID settings were used that allow convergence of temperature after the target temperature was set. For the mouse experiment, the mouse was 25 grams, therefore convergence onto the designated temperature (39 and 36 °C), occurred within ~20 minutes of setting the target temperature. Conversely, in the rat experiment, the rat weighed 801 grams and convergence onto the designated temperature took more than 100 minutes when the temperature difference was large. This larger time constant for convergence was mainly driven by the staggering weight difference between the rat and mouse at 32x; however, upon convergence, temperature was very stable for both species. See table SM2 for detailed summary of the results.

MR thermometry using magnetic resonance spectroscopy (MRS) require stringent magnetic field homogeneity conditions. Placement of optical temperature probes in the brain resulted in alteration of metabolite peak shapes, which made direct temperature estimation using MRS difficult, therefore, a separate experiment in which the brain and body temperature were modified over time to measure the time course of adaptation between the temperature of the body to that of the brain was conducted. Figure. 5 demonstrates the temperature difference between brain and body measured using optical probes placed in the brain and the rectum, demonstrating that it took less than 30 minutes for the temperature of the mouse to stabilize using the control system we developed. In minute ~60 of the experiment (after T1) the system was turned off to illustrate a reduction in body temperature of ~5 °C over the course of 40 minutes that occurred in the absence of heating. This reduction of body temperature was slower compared to the rapid cooling that was induced between T4 (minute ~205) and T5 (minute ~225), induced by the system. Body temperature overshoots of up to ~1.2°C were observed and those stabilized within <20 minutes. A summary of the results is reported in table SM3 in supplemental materials. Non-invasive MRS thermometry results are shown in Figure 6. The reconstructed temperature was within <0.3 °C of the target body temperature as measured by the thermistor placed in the rectum of the animal demonstrating the capability of the system to modulate brain temperature. Optical probe and spectroscopic temperature measurements were conducted independently to ensure correlation between the two as well as ensure that the placement of the optical probe in the brain did not alter thermoregulation.

## Discussions

In this work, we present an open-source system for controlling animal temperature using Peltier technology, allowing cooling and heating of rodents while mitigating the thermal regulation challenges that occur due to anesthesia. Heating or cooling of the Peltier modules was accomplished by switching the voltage polarity to the Peltier modules while varying the duty cycle of the Peltier modules currents controlled the power output. Heat exchange between the Peltier modules and water was conducted using a water heat exchanger and a pump that transferred the cold or hot water to the animal bed. The temperature of the animal was measured using a thermistor and fed back to an Arduino UNO based PID controller that was used to change the duty cycle and polarity of the Peltier modules to lock onto the target temperature. To the best of our knowledge, this is the first-of-its-kind open-source system capable of attaining control animal temperature under these conditions.

Since it is well known that temperature affects MR properties of tissues as well as animal physiology, locking the temperature following temperature modulation in either the positive or negative direction represented a highly desirable goal of this work. Particularly, it may impact the capability to conduct preclinical quantitative MRI, MR thermometry, fMRI, and other methods that are dependent on the MR properties and animal physiology. To this end, we demonstrated temperature control in a phantom where we showed that the system is capable of cooling a phantom to 15°C and heating it up to 45°C and that PID control improves temperature stability over time. With respect to the temperature probe design, our goal was to develop an inexpensive temperature probe (in contrast to optical systems that are proprietary and expensive) such that it does not cause large blooming artifacts due to susceptibility effects and does not heat up due to coupling with the transmit coil. In this respect, the thermistor used generated an artifact with a diameter of roughly 11mm, thus imaging objects >15mm away from the probe location did not induce any artifacts. Often, the desired imaging volume is more than 15 mm away from the rectum making the probe desirable for most imaging applications. RF interactions, which can be responsible for heating the animal and the probe, were mitigated by placement of 4 RF-chokes in close proximity to the thermistor and at subsequent λ/4 increments. No RF-heating was observed in all phantom and in vivo scans we conducted where temperature was tracked with both optical probes placed adjacent to the thermistor (SM Figure 1). The absence of interaction between the RF fields and the thermistor was also characterized in unperturbed $B_1^+$ and SNR maps (Figure 3 A and B).

In vivo temperature control was demonstrated in a mouse and a rat model, effectively locking onto 36 and 39 °C, respectively; however, given the large weight difference between the mouse and rat (32-fold difference), modification of the PID settings were necessary and the time for which the temperature converged was different for the two species, further demonstrating the flexibility and adaptability of the PID formalism for such biological experimentation. Further optimization of the PID settings would likely reduce the time for which it takes to lock onto the target temperature and reduce temperature oscillations over time, however, this can be a subject of future study. Similarly, changing the output power of the Peltier modules can further reduce the necessary time for heating or cooling the animal. One limitation of the system is the relatively long time required to transfer or dissipate heat from the animal. This limitation occurred due to the inefficient heat transfer of the system, namely, heating Peltier modules, then transferring (or taking away) heat to a circulating water bath that then interfaces with the animal through the water bed. While other heating devices (e.g., resistive heating) can possibly transfer heat more efficiently to the animal, to the best of our knowledge, there are no systems capable of cooling and heating without creating MR artifacts and having active feedback. Exploration of the use of highly thermally conductive solid materials (e.g., Aluminum Nitride), which are MR-invisible and can directly interface with the Peltier modules, may introduce more efficient ways to control the temperature of the animal. Rather than using water as a medium to control the bed temperature, direct thermal conduction from a Peltier module to a bed made of highly thermally conductive solid material, can be used to heat or cool the animal. Such systems however, would be markedly more expensive, compared to water-based transport systems. We chose to use water transport for this work as it offers a relatively inexpensive solution.

Implantation of the optical temperature probe in the brain resulted in a reduction of $B_0$ field homogeneity, which challenged MRS temperature measurement conducted by measuring the frequency shifts between water and NAA, Cho, and Cr, respectively. Additional experiments therefore were conducted to identify the time at which body temperature converges with brain temperature and we separately then conducted MRS experiments on an intact mouse, where the $B_0$ perturbation by the implanted brain probe was not present. This resulted in a non-invasively measured brain temperature that was <0.3°C from the rectal temperature measured using the thermistor, further proving modification of the brain temperature of the anesthetized animal and indicating this system to be a valuable tool for preclinical neuroscience.

**Acknowledgments**

This work was supported in part by NIH grant P41-EB017183.

# Figures

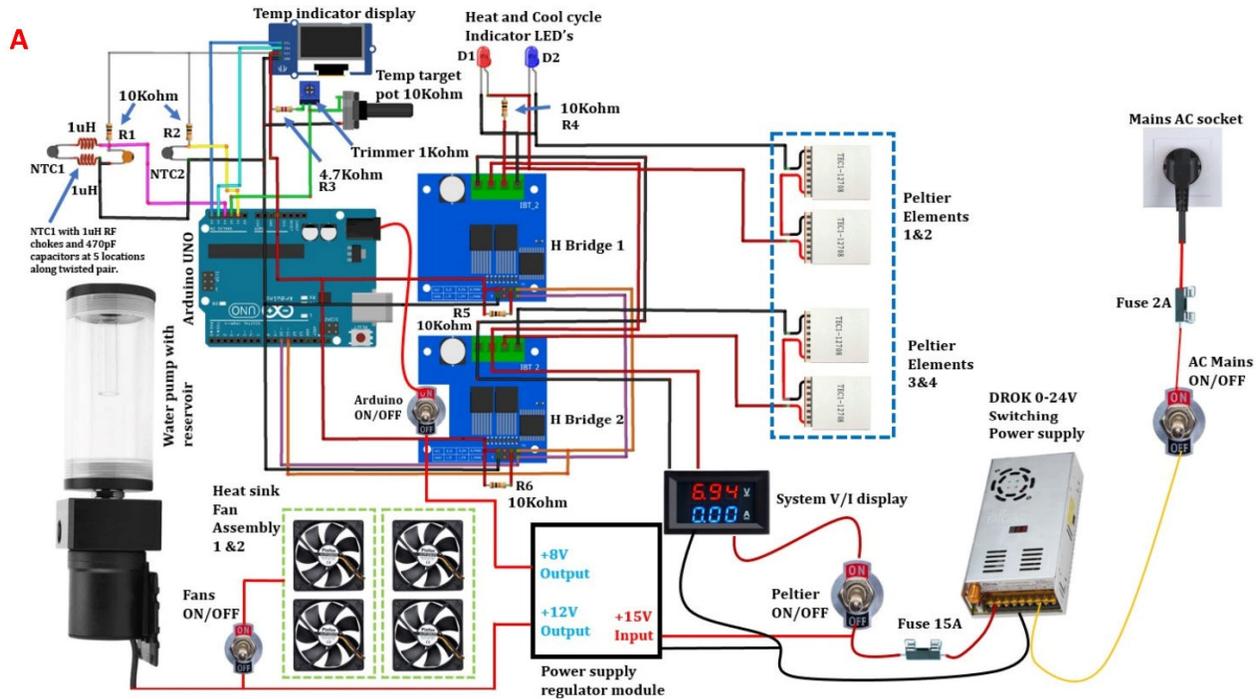

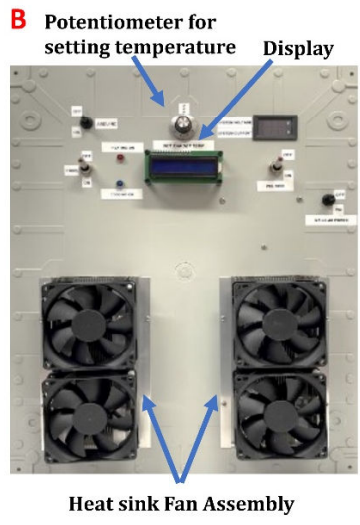
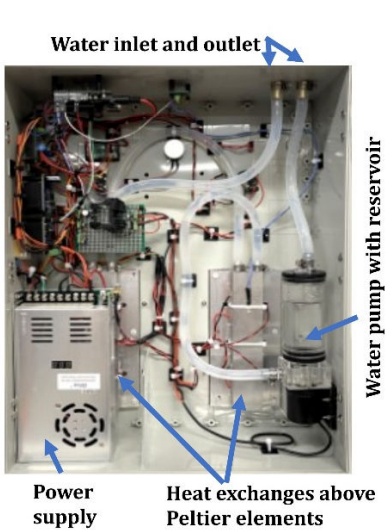

**Figure 1. A.** system architecture for the heating/cooling system and temperature measurement set-up. **B.** Front and back pictures of the designed box. **C.** 3D printed bed design with water channels allowing water to flow within the bed.

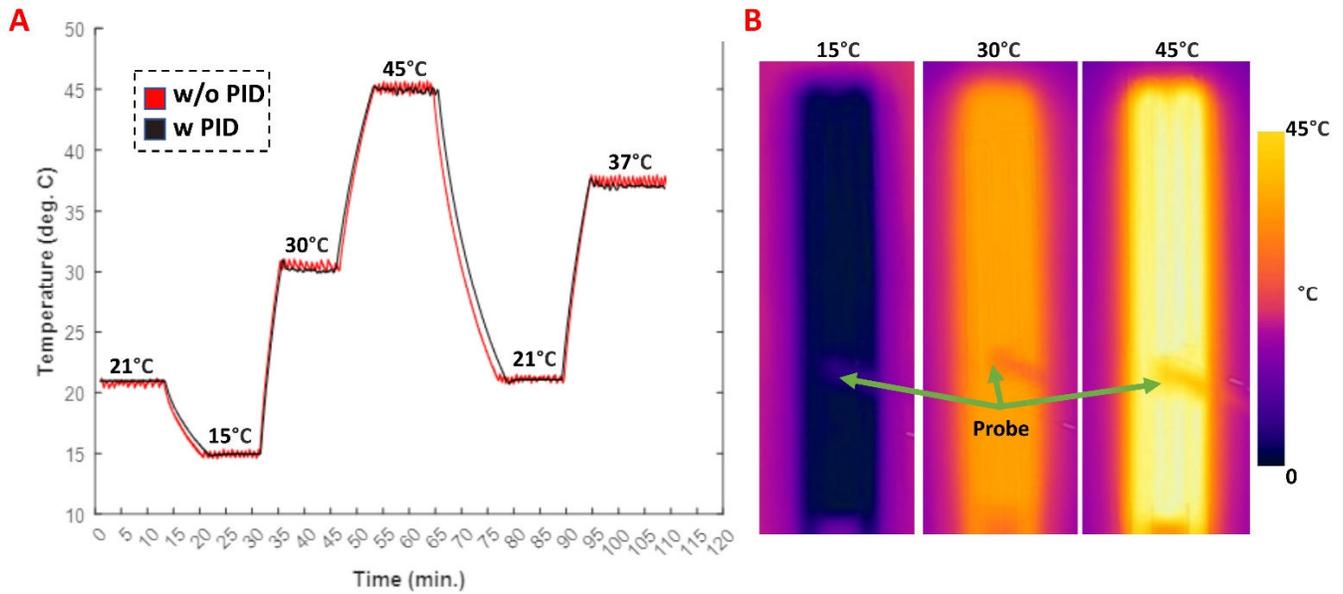

**Figure 2. A.** Comparison of temperature control with and without PID control in a liquid phantom. Reduction of temperature fluctuations was observed with the PID control. **B.** Infrared images of the animal bed at 15, 30 and 45 °C.

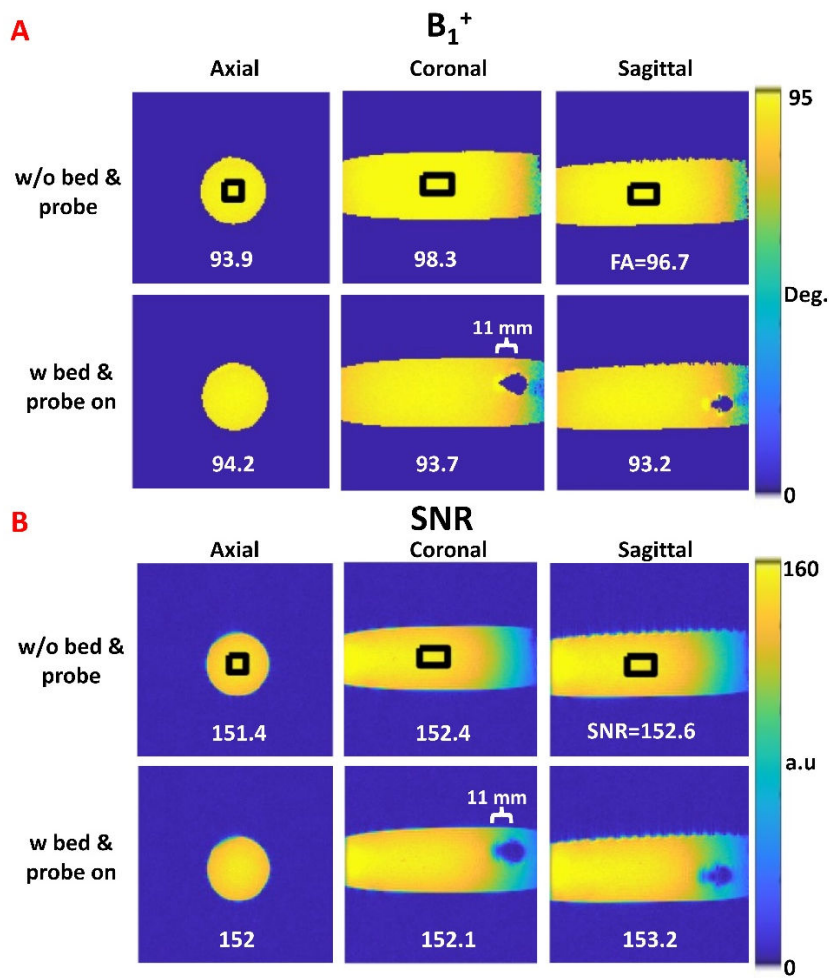

**Figure 3.** Comparison of $B_1^+$ (A) and SNR (B) maps for axial, coronal, and sagittal slices with and without the temperature probe inserted inside the phantom.

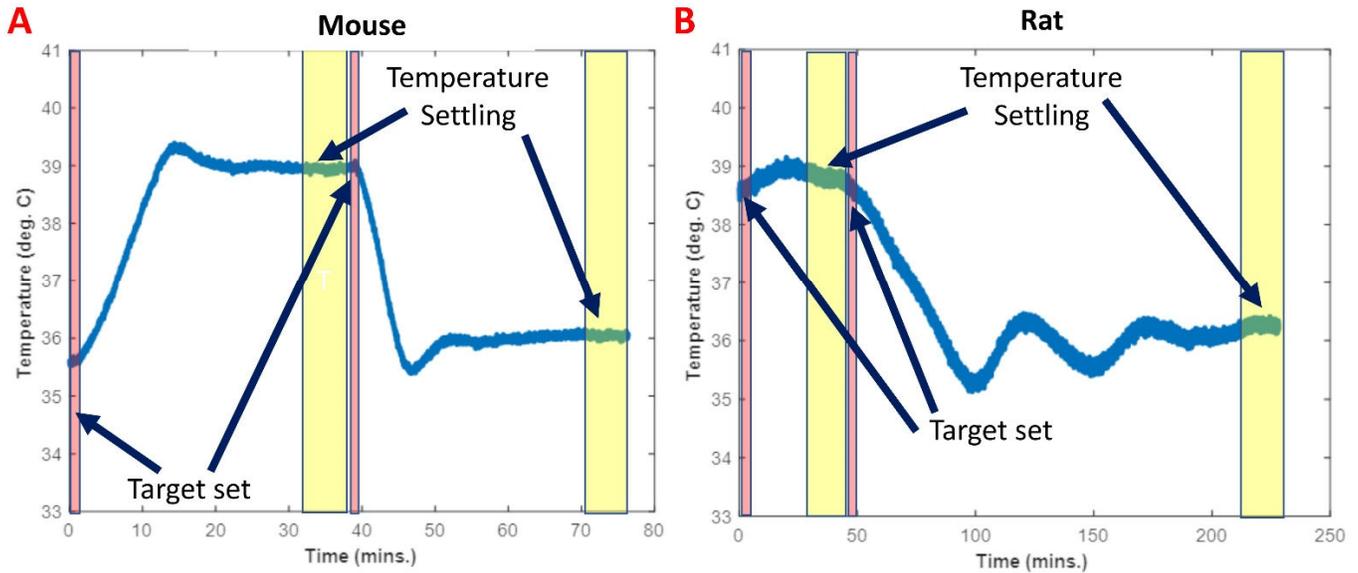

**Figure 4.** Time course of heating of the mouse (A) and rat (B) at 36 °C and 39 °C, respectively. Body temperature was measured using a fiber optic temperature probe placed in the rectum of the animal. Red blocks indicate the time at which the target temperatures were set and the yellow blocks show the time for which temperature stabilized.

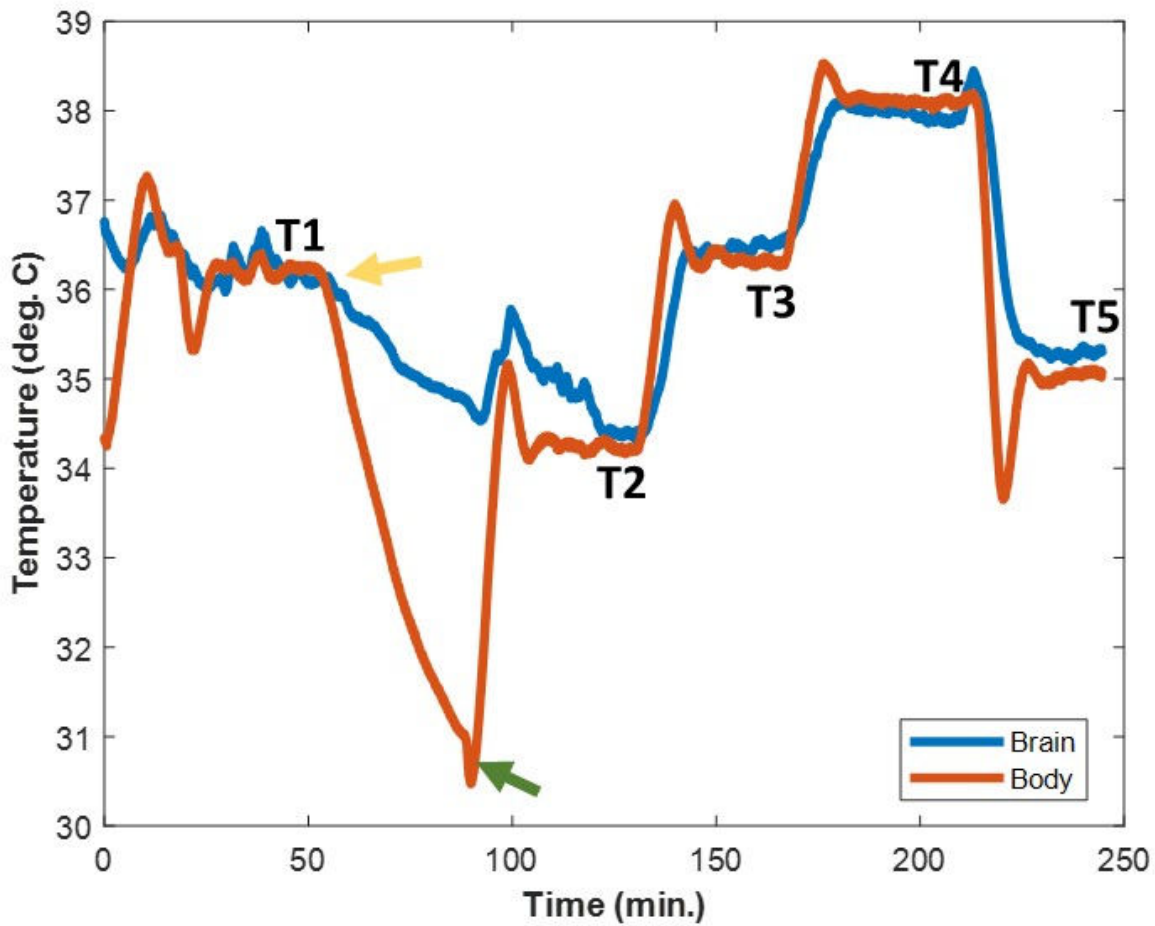

**Figure 5.** Comparison of body and brain temperature in an anesthetized mouse. T1-5 indicate the target temperatures at: 36, 34.1, 36.5, 38, and 35 °C, respectively. Note that the absent of heating

(minute ~60, yellow arrow), body temperature dropped rapidly, while when heating resumed (at minute 90, green arrow), temperature elevated by ~3.6 °C in 15 minutes.

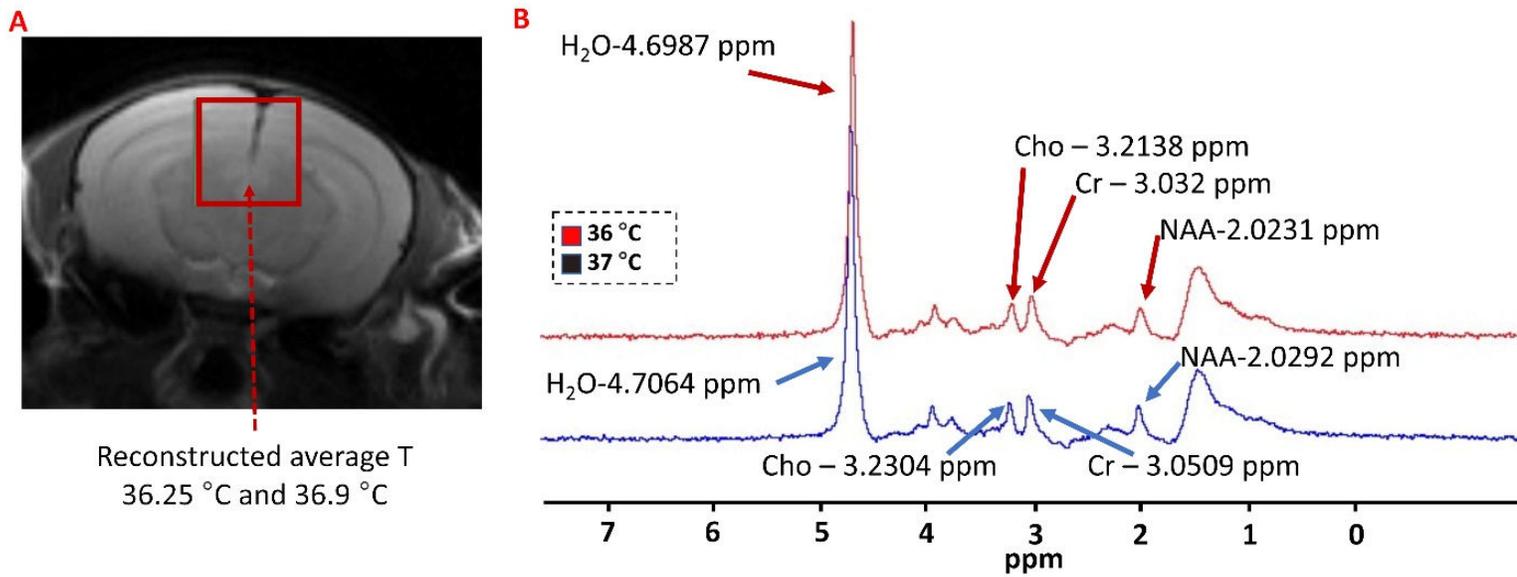

**Figure 6.** Mouse MRS thermometry at two different target temperatures of 36 and 37 °C respectively. Average temperature computed from NAA, Cr, and Cho distances to water was 36.25 and 36.9 °C, respectively. Mean $B_0$ drift of 0.014 was observed between the two experiments.

# Supplemental materials: Autonomous animal heating and cooling system for temperature-regulated MR experiments

| Designated T (deg. C) | PID | | | | | w/o PID | | | | | Swing improvement ratio |
|---|---|---|---|---|---|---|---|---|---|---|---|
| | Mean | Std | Max | Min | Max swing | Mean | Std | Min | Max | Max swing | |
| 21 | 20.98 | 0.02 | 21.06 | 20.87 | 0.19 | 20.80 | 0.18 | 21.19 | 20.35 | 0.84 | 4.42 |
| 15 | 14.96 | 0.02 | 15.05 | 14.9 | 0.15 | 14.96 | 0.17 | 15.36 | 14.59 | 0.77 | 5.13 |
| 30 | 30.03 | 0.042 | 30.17 | 29.92 | 0.25 | 30.49 | 0.23 | 31.05 | 30.03 | 1.02 | 4.08 |
| 45 | 44.89 | 0.17 | 45.18 | 44.46 | 0.72 | 45.13 | 0.28 | 45.72 | 44.46 | 1.26 | 1.75 |
| 21 | 21.09 | 0.02 | 21.15 | 21.02 | 0.13 | 21.15 | 0.15 | 21.50 | 20.77 | 0.73 | 5.62 |
| 37 | 37.02 | 0.04 | 37.14 | 36.92 | 0.22 | 37.42 | 0.27 | 37.98 | 36.90 | 1.08 | 4.91 |
| | | | | Mean: | 0.23 | | | | Mean: | 0.932 | 4.06 |

SM Table 1. Temperature stability with and without PID control in phantom.

| Designated T (deg. C) | Mouse | | | | | Rat | | | | |
|---|---|---|---|---|---|---|---|---|---|---|
| | Mean | Std | Max | Min | Max swing | Mean | Std | Min | Max | Max swing |
| 39 | 38.95 | 0.03 | 39.04 | 38.84 | 0.2 | 38.77 | 0.04 | 38.95 | 38.63 | 0.32 |
| 36 | 36.04 | 0.03 | 36.13 | 35.94 | 0.19 | 36.22 | 0.07 | 36.41 | 36.01 | 0.40 |

SM Table 2. Mouse and rat temperature control.

| Heating or cooling experiment | Designated T (deg. C) | Mean brain temperature (deg. C) | Mean body temperature (deg. C) | Mean Error (deg. C) |
|---|---|---|---|---|
| T1 | 36 | 36.14 | 36.24 | -0.10 |
| T2 | 34.1 | 34.39 | 34.21 | 0.18 |
| T3 | 36.5 | 36.53 | 36.31 | 0.22 |
| T4 | 38 | 37.97 | 38.11 | -0.13 |
| T5 | 35 | 35.31 | 35.07 | 0.23 |

SM Table 3. Mouse brain and body temperature across five different temperatures (T1-T5).

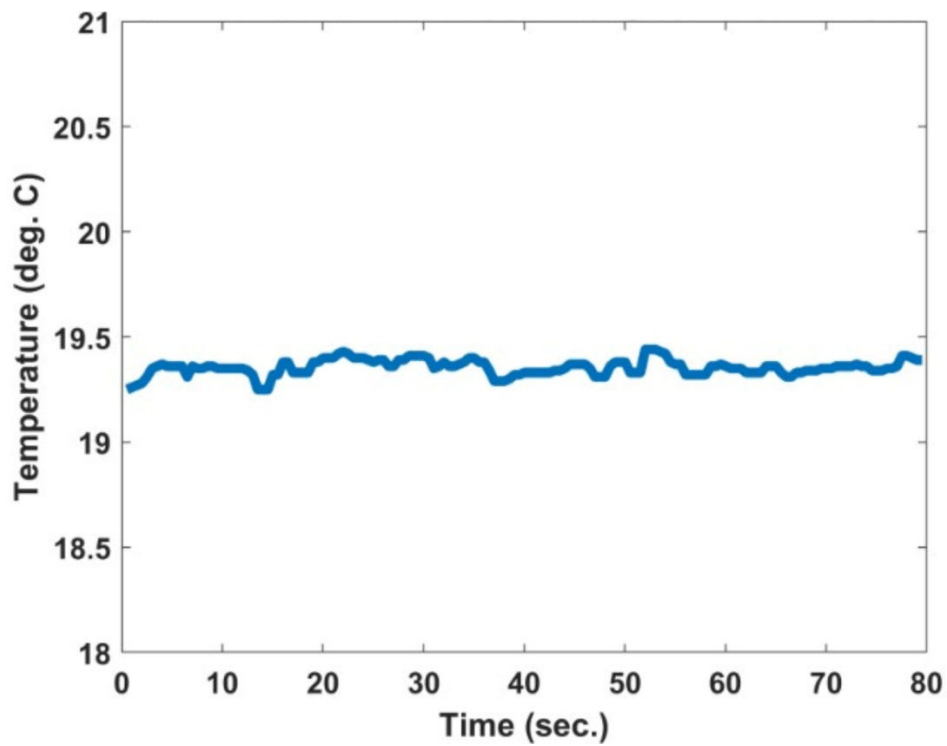

SM Figure 1. temperature during a high-SAR spin echo sequence. Currents were blocked using RF chokes on the cable of the thermocouple, resulting in no heating.

URL for design files and bill of materials can be found here:
https://www.dropbox.com/sh/o9y8k8m5s7cgswl/AADygTgGWvw5G0T-rpq9Ra9Ha?dl=0